\documentclass[twocolumn,aps,prb]{revtex4}

\usepackage{amsmath}
\usepackage{amstext}
\usepackage{amsfonts}
\usepackage{graphicx}
\usepackage{floatflt}
\usepackage{hyperref}
\hypersetup{colorlinks=true, linkcolor=blue, urlcolor=blue}
\usepackage{braket}
\usepackage{dcolumn}
\usepackage{bm}
\def\ie{\emph{i.e.},\ }

\def\ea{\emph{et~al.\ }}

\begin{document}
\title{Charge solitons and their dynamical mass in 1-D arrays of Josephson junctions}

\author{Jens Homfeld}
\affiliation{Institut f\"ur Theorie der Kondensierten Materie, Karlsruhe Institute of Technology, 
76128 Karlsruhe, Germany}

\author{Ivan Protopopov}
\affiliation{Institut f\"ur Theorie der Kondensierten Materie, Karlsruhe Institute of Technology, 
76128 Karlsruhe, Germany}
\affiliation{Landau Institute for Theoretical Physics, 119334 Moscow, Russia}

\author{Stephan Rachel}
\affiliation{Department of Physics, Yale University, New Haven, CT 06520, USA}

\author{Alexander Shnirman}
\affiliation{Institut f\"ur Theorie der Kondensierten Materie, Karlsruhe Institute of Technology, 
76128 Karlsruhe, Germany}
\affiliation{DFG Center for Functional Nanostructures (CFN), Karlsruhe Institute of Technology, 
76128 Karlsruhe, Germany}

\begin{abstract}
We investigate the charge transport in one-dimensional arrays of Josephson junctions.
In the interesting regime of "small charge solitons" (polarons), 
$\Lambda E_J > E_C > E_J$, where $\Lambda$ is the (electrostatic) screening length, 
the charge dynamics is strongly 
influenced by the polaronic effects, i.e., by dressing of a Cooper pair by charge dipoles.  
In particular, the soliton's mass in this regime scales approximately as $E_J^{-2}$. We employ 
two theoretical techniques: the many body tight-binding approach and the mean-field approach.
Results of the two approaches agree in the regime of "small charge solitons".
\end{abstract}

\maketitle

\section{\label{intro}Introduction}

Physics of one- and two-dimensional arrays of Josephson junctions is surprisingly 
rich. Both 1-D~\cite{PhysRevB.30.1138,JETPLett.60.738,PhysRevB.54.1228,PhysRevB.54.R6857,PhysRevB.54.1234,PhysRevLett.79.3736,JLTP.118.733,JLTP.124.291,JLTP.135.245} 
and 2-D~\cite{SovJETP.51.1015,PhysRevLett.65.645,PhysRevLett.65.923,PhysRevB.43.5307,PhysRep.355.235} arrays 
(including granulated superconducting films) have been extensively investigated. 
Yet, many unanswered questions remain. 
In particular, transport properties of 1-D arrays of Josephson junctions are still 
not fully understood. Experiments~\cite{PhysRevB.54.R6857,JLTP.118.733,JLTP.124.291} 
show various phenomena related to superconductor-insulator 
transitions, Coulomb blockade, hysteresis, mixed Josephson-quasi-particle effects etc..
One of the challenging questions is the value and the origin of the mass of the charge carriers
in the insulating regime. In the theoretical studies of Hermon \ea~\cite{PhysRevB.54.1234} it was shown 
that, if the grains have a large kinetic (or geometric) inductance, the system's dynamics
are governed by the sine-Gordon model and, therefore, kink-like
topological excitations, \ie charge solitons, are the charge carriers.
In Ref.~\onlinecite{JLTP.135.245} the 
domain of applicability of this sine-Gordon description was analyzed.
Simultaneous experiments by Haviland and Delsing~\cite{PhysRevB.54.R6857} 
demonstrated the Coulomb blockade in 1-D arrays of JJs consistent with the existence 
of charge solitons.  In the later experiments of Haviland's
group~\cite{JLTP.118.733,JLTP.124.291} considerable hysteresis in the $I$-$V$
characteristic of the array was observed and attributed to a very
large kinetic inductance. The physical origin of this inductance
remained unclear. A few years later, Zorin~\cite{PhysRevLett.96.167001} pointed out
that a current biased small-capacitance JJ develops an inductive
response on top of the capacitive one. This phenomenon was called {\it
Bloch inductance}. A closely related inductive coupling between two
charge qubits was studied in Ref.~\onlinecite{EPL.74.1088}. 
The role of the Bloch inductance in Josephson arrays was studied in Ref.~\onlinecite{PhysRevLett.96.167001} for the case of an infinite screening length, i.e., when the array 
serves as a zero-dimensional lumped circuit element. 

In this paper we employ two complimentary techniques to study the charge propagation
in infinite Josephson arrays with finite but large screening length. We consider arrays 
free of disorder. Specifically we concentrate on calculating the 
effective mass of the charge carriers. Both approaches, the many-body tight-binding 
technique and the mean-field technique agree for not very small ratios $E_{J}/E_C$. 
In particular, the effective mass of a charge soliton scales approximately as $E_J^{-2}$
in this regime. For full transport description one has to treat the effects of the array's 
boundaries as well as those of the disorder. Yet, our result about the effective mass
is clearly relevant for further investigation of the transport. 

\section{The system}

\begin{figure}[!htb]
\includegraphics[width=0.95\columnwidth]{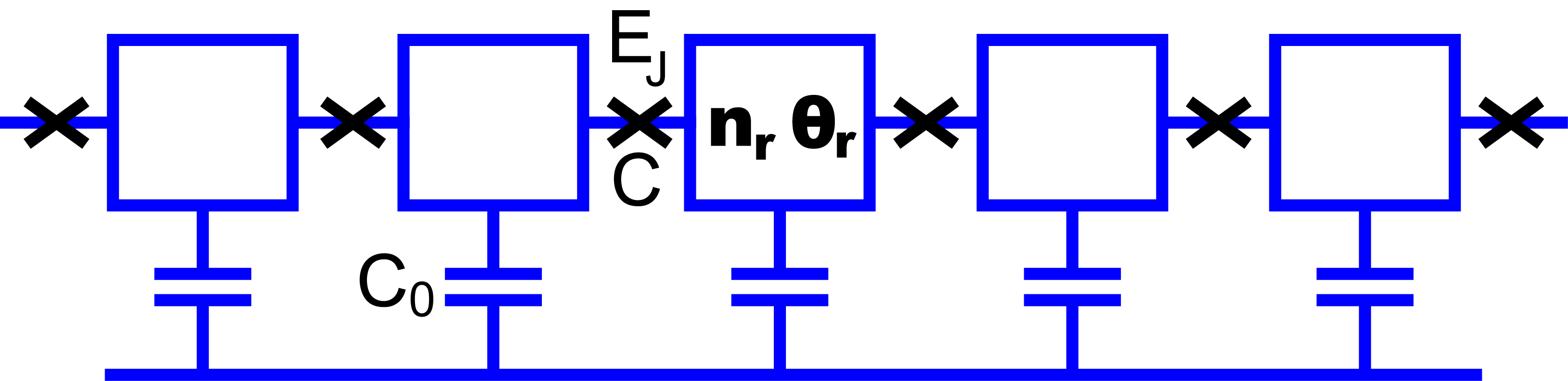}
\caption{Array of Josephson junctions.}
\label{fig:array}
\end{figure}

The system under study is shown in Fig.~\ref{fig:array}.
The Josephson junctions with capacitance $C$ connect the superconducting grains to each other and each grain has a capacitance $C_0$ to the ground. Typical values are $C\sim$ 1fF and $C_0\sim 5-20$ aF.
The system is governed by the usual Hamiltonian consisting of the Coulomb charging energy (kinetic energy)
and the Josephson tunneling (potential energy):  
\begin{equation}
	H=\frac12\sum_{r,r'}U(r-r')n_rn_{r'}-E_J \sum_{r}\cos \left(\theta_r-\theta_{r-1}\right)\ .
	\label{Eq:H_charge}
\end{equation}
Here $n_r$ are integer-valued island charges (in units of $2e$) and $\theta_r$ are the corresponding canonically conjugate phases, $\left[n_r,e^{i\theta_{r'}}\right]=e^{i\theta_r}\delta_{rr'}$. 
The matrix of Coulomb interaction $U(r)$ is given by
\begin{equation}
	U(r)=2E_C\int_{-\pi}^{\pi}\frac{dk}{2\pi}\frac{e^{ikr}}{\Lambda^{-2}-2\left( \cos k -1 \right)}\ ,
\end{equation}
with $E_C\equiv (2e^2)/2C$ being the charging energy and $\Lambda \equiv \sqrt{C/C_0}$ the screening length which determines the spatial extent of the Coulomb interaction. In this paper we consider $\Lambda \gg 1$.

\section{\label{sec:tb}Tight-binding approach}

\subsection{Qualitative discussion}
\label{subsec:qualit_disc}

In this section we explore the properties of Josephson arrays in the Coulomb blockade regime $E_J \ll E_C$. 
We consider the sector of the Hilbert space with exactly one extra Cooper pair in the array.  
The simplest (and having  minimal charging energy) 
representative of the unit charge sector is the state in which the extra Cooper pair resides on some island $R$ with 
all the other islands being neutral. The charging energy of such a state is given by $\mu_0 \equiv \frac12 U(0) \approx \Lambda E_C/2$. This is approximately the energy (rather high!) one has to 
invest to insert one Cooper pair into the array. Once the Cooper pair has been inserted it is free to move from one site to its neighbor via the 
hopping provided by the Josephson part of the Hamiltonian. In the limit of vanishingly small $E_J$ only the simplest charge configurations described above 
are important and  we are led to the trivial tight-binding  band $E(k)=-E_J \cos k$ for an extra Cooper pair in the Josephson chain (cf. \cite{JETPLett.60.738,PhysRevB.54.1228}).

The peculiarity of the $1D$-Josephson chain, first noticed in Refs.~\cite{JETPLett.60.738,PhysRevB.54.1228} and used in Ref.~\onlinecite{PhysRevB.80.180508}, is that  the simple picture sketched above is valid only for 
extremely small $E_J< E_C/\Lambda$. The reason is the presence of a large number of states lying at small energy $\sim E_C/\Lambda$ above the basic
states (as opposed to much larger energy $E_C$ which one might expect and which indeed happens in higher dimensions). 
One particular example is the charge configuration $|1, -1, 1\rangle $ (Cooper pair and a properly oriented dipole nearby) having the energy
$\frac32 U(0)-2U(1)+U(2)\approx \mu_0+E_C/\Lambda$. 
Thus, in the parameter range $\Lambda E_J > E_C > E_J$ called by the authors of 
Ref.~\onlinecite{PhysRevB.80.180508} the small soliton regime, Cooper pair inserted into 
the chain gets strongly dressed by virtual dipoles and the simplest tight-binding scheme breaks down.
Dipole dressing was also mentioned in the context of transport in ion channels~\cite{PhysRevLett.95.148101}.  

In reference \cite{PhysRevB.80.180508} the properties of the small charge solitons were addressed  by successive inclusion of the charge configurations
(up to $32$ states) with larger and larger energies into the tight-binding scheme. A similar scheme was developed for polarons in Ref.~\onlinecite{PhysRevB.60.1633}.
In this paper we construct a comprehensive description of the low-lying (with energies much smaller $E_C$) states in terms of a particular 
spin-$1/2$ model. We derive an effective Hamiltonian governing the model dynamics within the low energy subspace. We then develop a tight-binding approach 
with arbitrary number of the charge states taken into account. 

\subsection{Structure of the low energy subspace}
\label{sec:Structure}

Let us first define more precisely what we mean under the low lying states in the sector with total charge $1$  and construct the complete classification 
of these states. Let us consider some charge configuration of size $w$. It is clear that the energy of such a configuration will certainly exceed $E_C$ if 
$w> \Lambda$ (from now on we count energies from the energy $\mu_0$ of a single  Cooper pair). Thus for the low-lying configurations $w<\Lambda$ and we 
can expand the charging energy in powers of $1/\Lambda$ as 
\begin{equation}
	H_C=-\frac{E_C}{2}\sum_{rr'}|r-r'|n_rn_{r'} +\frac{E_C}{4\Lambda}\sum_{r, r'}(r-r')^2 n_rn_{r'}+\ldots
	\label{expanded_Ham}
\end{equation}
We see that the typical  charge configurations  have large energy $\sim  E_C\gg E_J$ and are not important for the low energy physics. The exceptions are the states nullifying 
the first term in Eq.~(\ref{expanded_Ham}) and having the energy $\sim w E_C/\Lambda$. 
Note, that the first term of the expansion (\ref{expanded_Ham}) can not take negative values. Otherwise there would exist configurations with electrostatic energy smaller 
than $\mu_0$. As long as $w < \Lambda E_J/E_C$ these charge configurations hybridize effectively with the basic one leading to the formation of the small charge soliton.
Thus, the condition 
\begin{equation}
	\sum_{rr'}|r-r'|n_rn_{r'}=0
	\label{Eq:Low_energy_space}
\end{equation}
is the mathematical definition of the low energy subspace in the unit charge sector (we call it also the proper space). 
It can be shown (Appendix \ref{App:Low_energy_space}) that  the subspace (\ref{Eq:Low_energy_space}) consists of all the configurations with two properties:
a) all the islands' charges $n_r$ equal $\pm 1$ or $0$; b) any two charged islands 
separated by an arbitrary number of neutral islands have opposite charges.
For example,  the configurations $|1, 0, -1, 1\rangle$ and $|1, 0, -1, 0, 1\rangle$ belong to the low-energy space while the configurations
$|2, 0, -1\rangle$ and $|1, 0, 1, 0, -1\rangle$ do not. 
To describe the proper subspace in a more clear way let us introduce variables 
$\sigma_r$ defined on the links of the chain (the link $r$ is the 
link connecting islands $r$ and $r+1$)
\begin{equation}
	\sigma_r^z=-\sum_{r'\geq r+1}n_{r'}+\sum_{r'\leq r}n_{r'}\quad,\quad n_r=\frac12\left( \sigma_r^z-\sigma_{r-1}^z \right)\ .
\end{equation}
\begin{figure}
	\includegraphics[width=0.95\columnwidth]{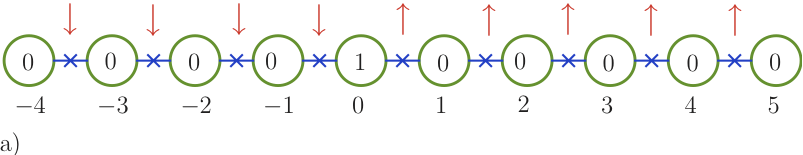}\\
	\vspace{0.5cm}
	\includegraphics[width=0.95\columnwidth]{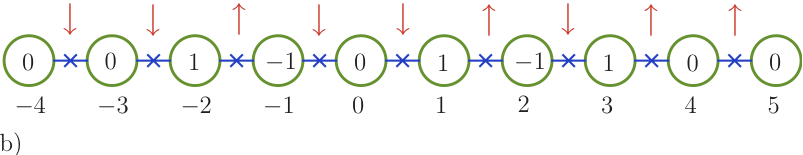}
	\caption{\small Connection between $\sigma^z_r$ (defined on the links) and the charges of the islands $n_r$. 
	In the low-energy subspace $\sigma^z_r$ can take only two values $\pm 1$ and are $z$-projections of a spin-$1/2$. 
	a) The basic charge configuration with only one charged island corresponds to an abrupt domain wall in terms of $\sigma_r^z$. 
	b) More complicated charge configuration (two additional dipoles) corresponding to a domain wall of finite thickness in the spin language. }
	\label{Fig:n_to_sigma}
\end{figure}

The connection between variables $\sigma_r^z$ and charges $n_r$ for two 
configurations in the low-energy subspace is illustrated in Fig.~\ref{Fig:n_to_sigma}.
From the definition of $\sigma_r^z$ and the properties of the states in the low-energy subspace one immediately concludes that the low-energy 
configurations are described by $\sigma_r^z=\pm 1$ for all $r$, i.e., the low-energy subspace is isomorphic to the space of states for a spin-$1/2$ chain
with $\sigma_r^z$ being the z-projections of the spins. 

Due to the constraint $\sum_{r} n_r=1$ the variable $\sigma^z_r$ satisfies the boundary conditions
\begin{equation}
	\sigma^z_{r\rightarrow-\infty}=-1\quad,\quad  \sigma^z_{r\rightarrow+\infty}=1\ .
	\label{Eq:cond}
\end{equation}
Thus, extra Cooper pair in the chain is described by a domain wall in the spin language.  

\subsection{Projecting the Hamiltonian}
\label{projham}

Having understood the structure of the low-energy space of the model we can project the full Hamiltonian (\ref{Eq:H_charge}) onto the proper subspace. 
The projection is carried out by noting that Cooper pair tunneling between two neighboring islands corresponds to the spin flip in the link between them. 
Thus the Josephson part of the Hamiltonian is given by
\begin{equation}
	H_J=-E_J \sum_{r}\sigma_r^x\ .
\end{equation}
Rewriting the charging energy in terms of spin variables we arrive at
\begin{equation}
	H=\frac{1}{8}\sum_{r, r'}(\sigma_{r}^z-\sigma_{r-1}^z)(\sigma_{r'}^z-\sigma_{r'-1}^z)U(r-r')-E_J\sum_{r}\sigma_r^x\ .
	\label{Eq:H_spin}
\end{equation}
To determine the spectrum of the single-charge sector of the Hamiltonian (\ref{Eq:H_spin}) 
we impose the boundary conditions (\ref{Eq:cond}) indicating the 
presence of a domain wall. 

The Hamiltonian (\ref{Eq:H_spin}) takes into account the low energy charge configurations
of arbitrary width $w$. We understand however that the configurations
with $w\gg\Lambda E_J/E_C$ are not important at low energies. Thus, we can further reduce the phase space by dropping out all the configurations of the 
width $w$ larger than some $w_0$. We expect that at $w_0\gg\Lambda E_J/E_C$ the resulting  low energy states are independent of $w_0$ and approximate correctly those of
Hamiltonian (\ref{Eq:H_spin}).

Any state containing a domain wall of the width less than $w_0$ is completely specified by the position $R$ of the first spin up
(which we call the coordinate of the charge soliton or domain wall) and the values of the $z$-projections of the next $w_0$ spins 
$\left\{\tilde{\sigma}_1, \ldots \tilde{\sigma}_{w_0}\right\}$. 
Given the state 
\begin{equation}
	|R\rangle|\tilde{\sigma}_1\,, \ldots \tilde{\sigma}_{w_0}\rangle
	\label{Eq:basic_state}
\end{equation}
one can reconstruct the $z$-projections of all spins in the chain according to
\begin{eqnarray}
	\sigma_r^z=\left\{
	\begin{array}{cc}
		-1\,, & r<R\ ,\\
		1\,, & r=R\ ,\\
		\tilde{\sigma}_{r-R}\,, & R+1\leq r\leq R+w_0\ ,\\
		1\,, & r>R+w_0\ .
	\end{array}
	\right.
\end{eqnarray}
For example, if we choose $w_0=5$ the states shown on Fig. \ref{Fig:n_to_sigma}a) and
\ref{Fig:n_to_sigma}b) can be written as
\begin{eqnarray}
	|R=0\rangle |\uparrow, \uparrow, \uparrow, \uparrow, \uparrow\rangle\ ,\\
	|R=-2\rangle |\downarrow, \downarrow, \uparrow, \downarrow, \uparrow\rangle\ .
\end{eqnarray}

In Appendix~\ref{App:Projection} we describe how to project the Hamiltonian (\ref{Eq:H_spin}) onto the space of configurations with sizes less or equal than $w_0$. We also perform a transition from the coordinates 
$|R\rangle$ to the quasi-momentum $k$. The result reads
\begin{multline}
	H=H_C -E_J  \sum_{r=1}^{w_0}\sigma_r^x\\-E_J\left[\sum_{r=1}^{w_0}
	e^{irk}	\left(T^\dag\sigma_1^+\right)^r\sigma_r^-
	+ e^{i\left( w_0+1 \right)k}\prod_{r=1}^{w_0}\sigma_r^++h.c.\right]\ ,
	\label{Eq:H_final}
\end{multline}
where $T$ is the operator of the right cyclic shift defined by
$T|\tilde{\sigma}_1\,, \ldots \tilde{\sigma}_{w_0}\rangle=|\tilde{\sigma}_{w_0}\,, \tilde{\sigma}_{1}\,,\ldots \tilde{\sigma}_2\rangle$.

The Hamiltonian (\ref{Eq:H_final}) constitutes the main result of this section. For $w_0=1, 2, \ldots 5$ it can be shown to produce results equivalent to that of reference \cite{PhysRevB.80.180508}. Equation (\ref{Eq:H_final}) reduces the initial many-body problem to a finite dimensional Hamiltonian, readily 
accessible to numerics as long as not too large ($w_0\leq 20$) charge configurations are important. 
In the next sections we present the results of numerical analysis of 
the Hamiltonian (\ref{Eq:H_final}) and compare the results to those of the mean-field approach.

\subsection{Results of the tight-binding approach}

\begin{figure}[!htb]
 \includegraphics[width=0.95\columnwidth]{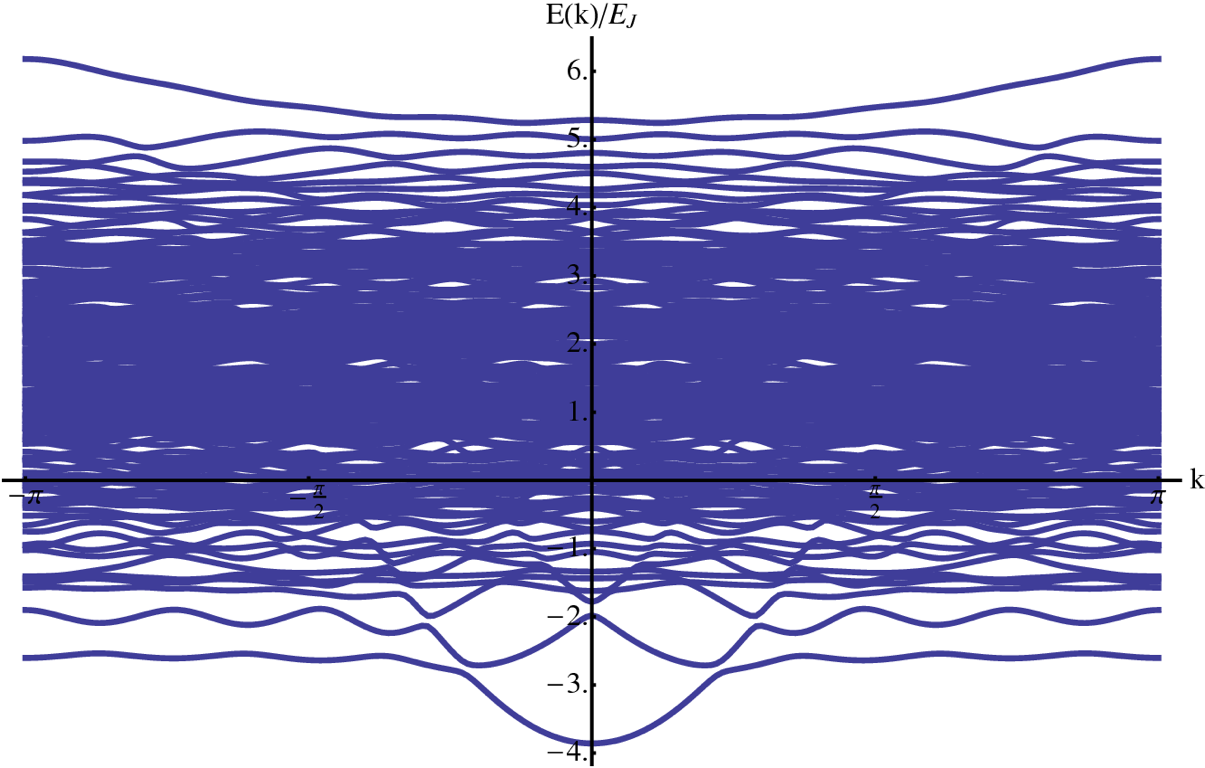}
\caption{Spectrum of the tight-binding Hamiltonian (\ref{Eq:H_final}) for $E_J/E_C = 0.4$ and $\Lambda=10$. The number of charge states taken into account equals $2^7$.}
\label{fig:7_alle}
\end{figure}

\begin{figure}[!htb]
\includegraphics[width=0.75\columnwidth]{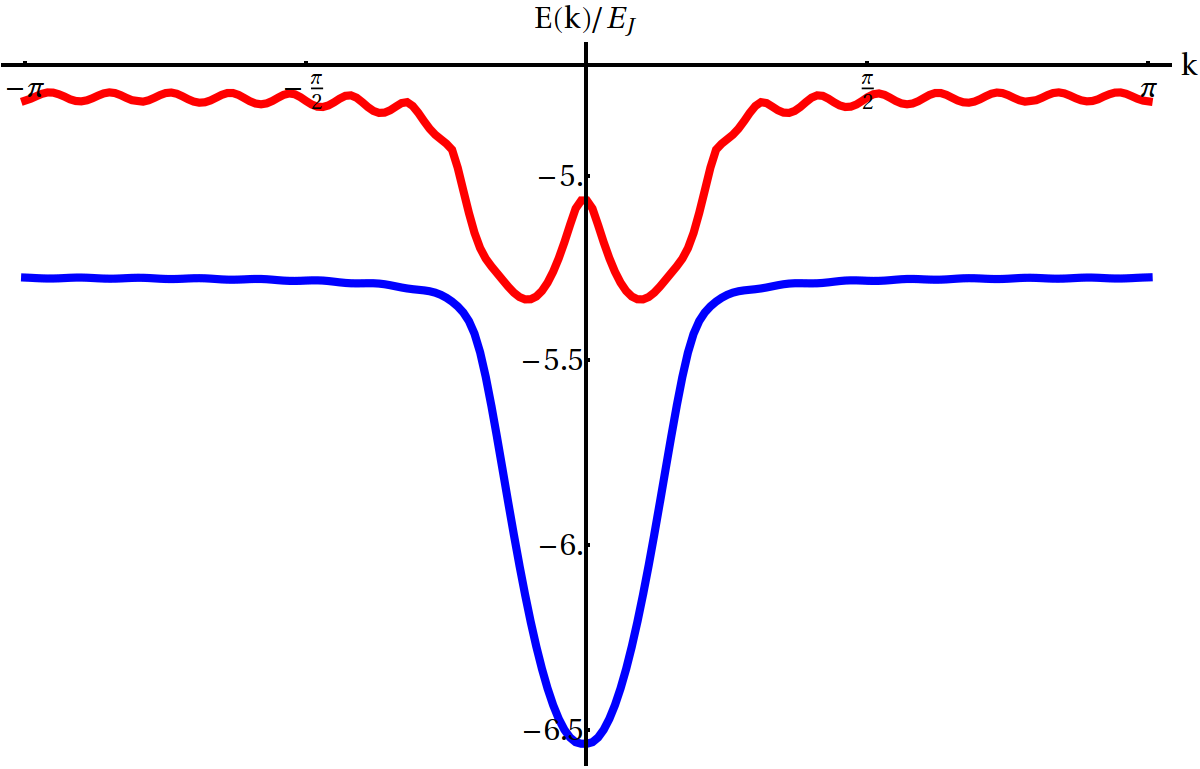}
\caption{Two lowest energy bands for $\Lambda=10$ and $E_J/E_C=0.4$. 
The number of charge states taken into account equals $2^{18}$.}
\label{fig:2bands}
\end{figure}

\begin{figure}[!htb]
\includegraphics[width=0.75\columnwidth]{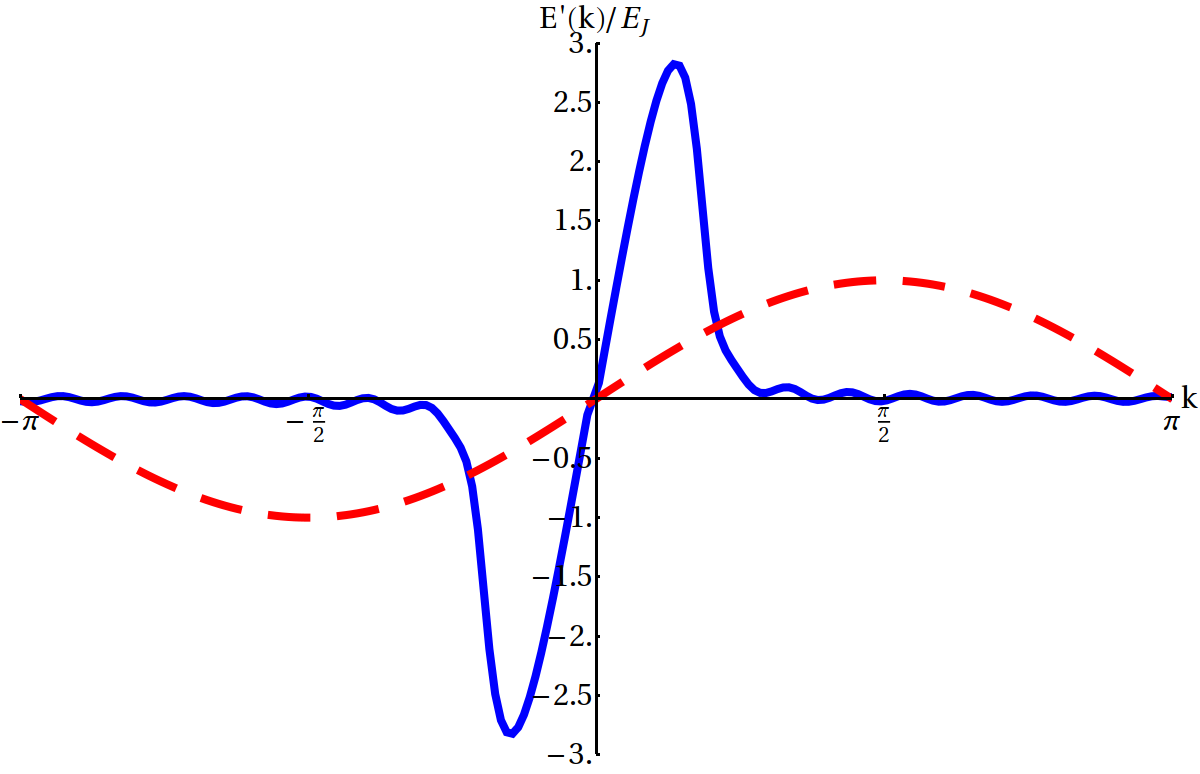}
\caption{Solid (blue) line: the group velocity corresponding to the lowest energy band of Fig~\ref{fig:2bands}.
Dashed (red) line: the naive tight-binding group velocity with no dressing by dipoles taken into account.}
\label{fig:GroupV}
\end{figure}

An example of the band structure obtained within the tight-binding approach is shown 
in Fig.~\ref{fig:7_alle}. In Fig.~\ref{fig:2bands} the two lowest bands are shown.
We observe that the lowest band is parabolic for small momenta $k$ and flattens in the 
outer part of the Brillouin zone. This phenomenon was already observed in Ref.~\onlinecite{PhysRevB.80.180508}. 
To further emphasize the dispersion relation of the lowest band in Fig.~\ref{fig:GroupV} we show the group velocity of the soliton (dressed Cooper pair) as compared to the one of an undressed Cooper pair. We find that the flattening of the dispersion relation in the outer region of the Brillouin zone leads to zero group velocity.
In this paper we concentrate mostly on the investigation of the effective mass of the charge carriers. In the tight-binding approach we define 
$m_{\rm TB} =\hbar^2\, \left(\frac{\partial^2 E_0(k)}{\partial k^2}\Big|_{k=0}\right)^{-1}$, 
where $E_0(k)$ is the dispersion of the lowest band (ground state). In what follows we will 
compare this mass with the results of the mean-field theory. 

\subsubsection{Persistent current}

As a first obvious application of our results consider a ring-shaped array of $N$ junctions 
with exactly one extra Cooper pair in it. If an external magnetic flux $\Phi_{ext}$ is applied a 
persistent current will emerge. The periodic boundary condition for the Bloch wave with 
wave vector $k$ reads
\begin{equation}
e^{ikN}=e^{2\pi \frac{\Phi_{ext}}{\Phi_0}}\ .
\end{equation} 
Thus, as the external flux varies between $-\Phi_0/2$ and $\Phi_0/2$, the relevant wave vector 
varies between $-\pi/N$ and $\pi/N$. For large enough $N$ the interval $[-\pi/N,\pi/N]$ is 
safely within the domain of parabolic dispersion relation. Thus we use the effective mass 
approximation and obtain for the persistent current in the interval 
$\Phi_{ext} \in [-\Phi_0/2, \Phi_0/2]$
\begin{equation}
I(\Phi_{ext}) \approx \frac{2e}{N} \,\frac{\hbar k}{m_{eff}} =\frac{2e}{N^2 m_{eff}} \,\frac{2\pi \hbar \Phi_{ext}}{\Phi_0} \ ,
\end{equation} 
where $m_{eff}$ is the effective 
mass of the charge carrier (in the tight-binding approach we 
obtained $m_{eff}=m_{\rm TB}$). Thus, the amplitude of the persistent current 
oscillations is given by 
\begin{equation}
I_0=\frac{2\pi\hbar e}{N^2 m_{eff}}\ .
\end{equation}
With no polaronic effects taken into account, i.e., for a bare Cooper pair we 
would have $E^{bare}_0(k) = -E_J \cos k$ and 
$m^{bare}_{eff}=\hbar^2/E_J$. Thus we obtain
\begin{equation}
I_0=\frac{2\pi e E_J}{\hbar N^2}\frac{m^{bare}_{eff}}{m_{eff}} 
= \frac{\pi I_c}{N^2}\frac{m^{bare}_{eff}}{m_{eff}} \ ,
\end{equation}
where $I_c$ is the critical current of a single Josephson junction.
We observe that the effective mass reduction via the polaronic effects 
enhances the persistent current.

\section{\label{sec:mf}Mean-Field Theory}

\subsection{Description in terms of continuous polarization charges}

\begin{figure}[!htb]
 \includegraphics[width=0.9\columnwidth]{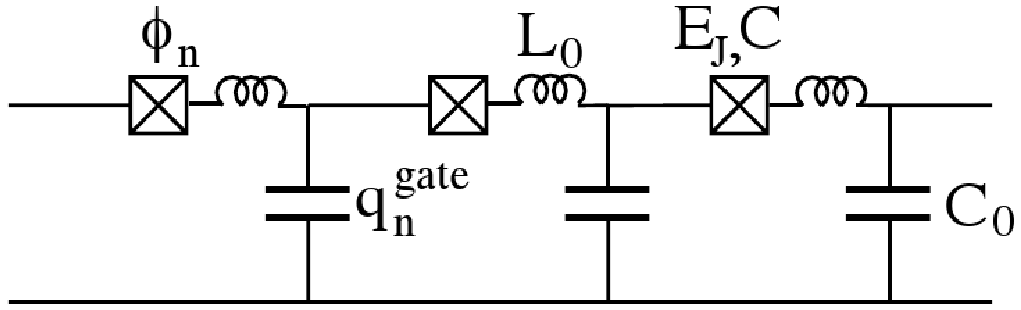}
\caption{Array of Josephson junctions with infinitesimal inductances $L_0$.} 
\label{elecscheme}
\end{figure}

An alternative description of the charge propagation in the array is given in terms of the 
continuous polarization charges, e.g., the screening charges $q_n^{\rm gate}$ 
on the gate capacitances $C_0$ (see Fig.~\ref{elecscheme}). 
For the system described in the previous section the 
continuous polarization charges are enslaved to the discrete charges $n_r$. That is, once 
a tunneling process occurs and the distribution $n_r$ changes, the polarization charges 
adjust immediately to the new situation. 
To allow formally independent dynamics of polarization 
charges we introduce infinitesimal inductances $L_0$ as shown in Fig.~(\ref{elecscheme}). 
This leads to two independent degrees of freedom per cell of the array. One quantized 
charge degree of freedom $m_r=\sum_{k=r}^{\infty}n_k$ is the number of Cooper pairs that have tunneled 
through junction number $r$. Its conjugate phase is given by $\phi_r=\theta_r-\theta_{r-1}$ and the 
commutation relations read $\left[m_r,e^{i\phi_{r'}}\right]=e^{i\phi_r}\delta_{rr'}$. The second continuous 
charge degree of freedom $Q_r \equiv\sum\limits_{r'<r}q_{r'}^{gate}+2em_{-\infty}$ is equal to the polarization 
charge that has arrived at the junction number $r$ or, alternatively, the integral of the displacement current 
flowing into junction $r$. The conjugate variable $\Phi_r$ is the magnetic flux on inductance $L_0$ in cell number $r$. The commutation relation reads $\left[\Phi_r,Q_{r'}\right]=i\hbar\delta_{rr'}$.
We obtain the following Hamiltonian of the array
\begin{eqnarray}
H&=&\sum\limits_r\left[\frac{(2em_r-Q_r)^2}{2C}-E_J\cos(\phi_r)\right.\nonumber\\&+&\left.\frac{(Q_r-Q_{r-1})^2}{2C_0}+\frac{\Phi_r^2}{2L_0}\right]\ .
\label{mfnham1}
\end{eqnarray}

\subsection{Mean-field approximation}

The mean-field description is based on the Heisenberg equations of motion for the polarization 
charge $Q_r$ following from (\ref{mfnham1}):
\begin{equation}
 L_0\ddot Q_r=-\frac{1}{C}(Q_r-2em_r)-\frac{2Q_r-Q_{r-1}-Q_{r+1}}{C_0}\ .
\label{heisenb}
\end{equation}
We average Eq.~(\ref{heisenb}) over the state of the system and obtain
\begin{equation}
L_0\langle\ddot Q_r\rangle=-V_r-\frac{2\langle Q_r\rangle-\langle Q_{r+1}\rangle-\langle Q_{r-1}\rangle}{C_0}\ ,
\label{skb1}
\end{equation} 
where $V_r\equiv\left\langle\frac{1}{C}(Q_r-2em_r)\right\rangle$ is the expectation value of the 
voltage drop across junction number $r$. In the mean-field approximation we 
calculate $V_r$ by replacing the operators $Q_r$ by their average values $\langle Q_r\rangle(t)$ in 
the Hamiltonian (\ref{mfnham1}).
Thus the problem factorizes to many single-junction ones. Each junction is governed by the 
Hamiltonian 
\begin{equation}
H(Q(t))=\frac{(2em-Q(t))^2}{2C}-E_J\cos\phi\ ,
\label{mfnham2}
\end{equation}
where we have dropped the index $r$. The gate charge 
$Q(t)$ is a given function of time (to be replaced in each junction by $\langle Q_r\rangle(t)$).
For the expectation value of the voltage we then obtain $V= \langle \partial_Q H\rangle$.
The problem is now to  find the quantum state of the junction in which the average 
$\langle \partial_Q H\rangle$ should be evaluated. We do so assuming that $\langle Q_r\rangle(t)$
is a slow function of time. This assumption should be checked for self-consistency later.

\begin{figure}[!htb]
 \includegraphics[width=0.9\columnwidth]{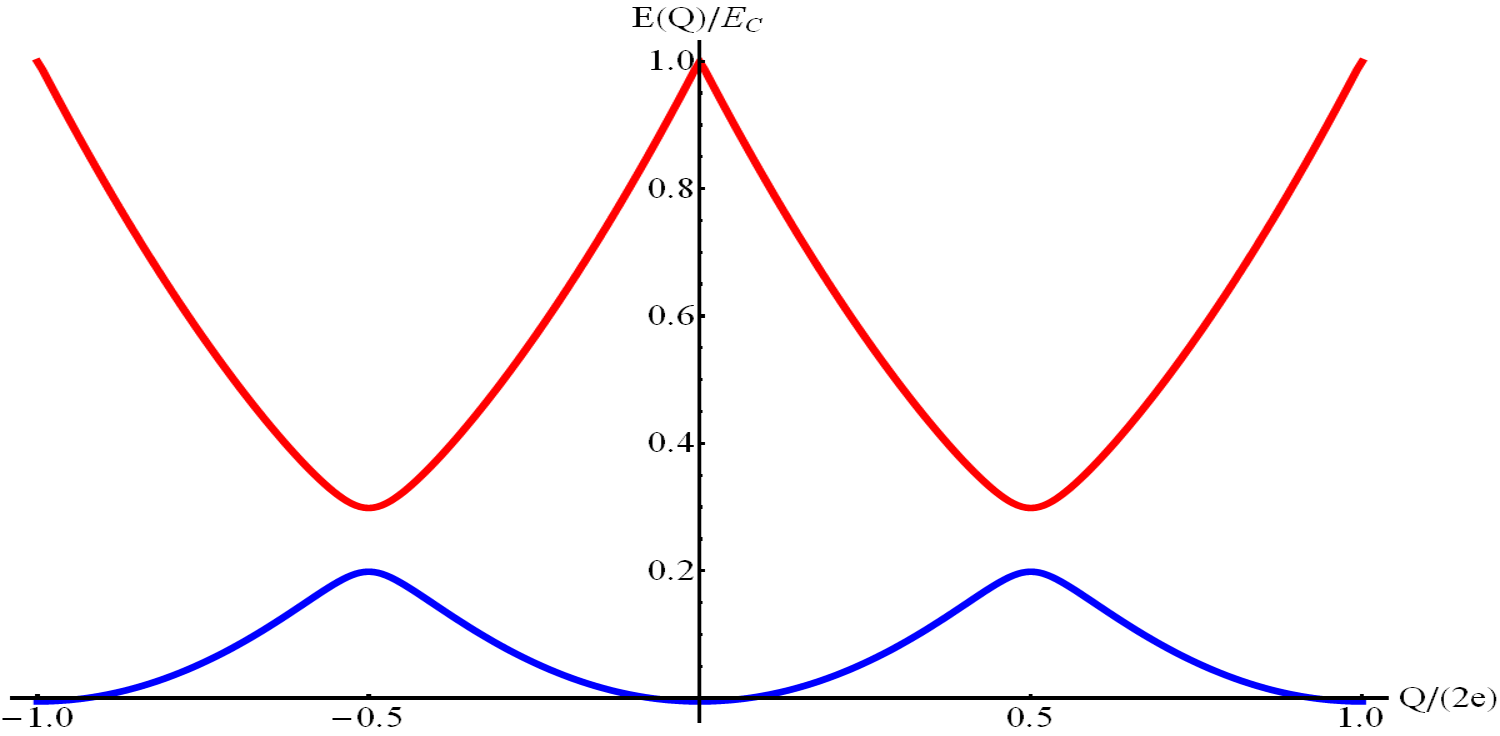}
\caption{The ground and first excited state $E_0,E_1$ of (\ref{mfnham2})}
\label{ew12}
\end{figure}

\begin{figure}[!htb]
 \includegraphics[scale=0.19]{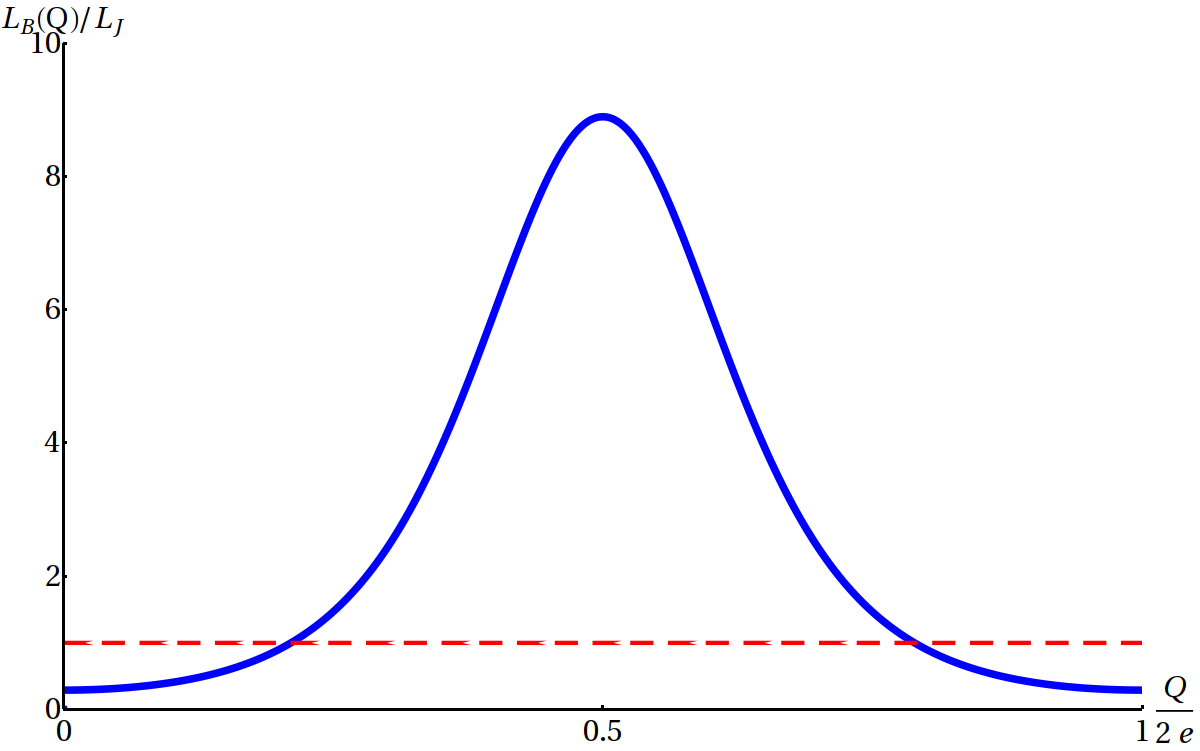}
\caption{Solid line: the Bloch inductance $L_B(Q)$ for $E_J/E_C=0.5$, measured in units of the Josephson inductance $L_J$. Dashed line: $L_B(Q)=L_J$ for $E_J/E_C = \infty$.}
\label{fig:BlochInductance}
\end{figure}

The Hamiltonian (\ref{mfnham2}) possesses the (adiabatic) spectrum 
with discrete eigenvectors $\Ket{e_n(Q(t))}$ obeying $\Braket{e_n|e_m}=\delta_{nm}$ and 
eigenvalues $E_n(Q(t))$, cf. Fig.~\ref{ew12}. 
The general wave function is a superposition $\Ket{\Psi(t)}=\sum\limits_n\alpha_n(t)\Ket{e_n(Q(t))}$.
Our aim is to determine $\Ket{\Psi(t)}$ for a given function $Q(t)$. We restrict ourselves to the adiabatic 
case, i.e., we keep only terms of order $\ddot Q, \dot Q^2$.
After a calculation presented in Appendix~\ref{App:Adiabatic} we arrive at
\begin{equation}
V=\langle \partial_Q H\rangle=\partial_Q E_0+L_B\ddot Q+\frac{1}{2}(\partial_Q L_B)\dot Q^2\ .
\label{spann}
\end{equation}
Here we defined the Bloch inductance first introduced by Zorin~\cite{PhysRevLett.96.167001}:
\begin{equation}
L_B=2\hbar^2\sum_{n>0}\frac{\Bra{e_n}\partial_Q H\Ket{e_0}^2}{(E_n-E_0)^3}\ .
\label{bloch}
\end{equation}
(In Ref.~\onlinecite{PhysRevLett.96.167001} only the first excited state ($n=1$) in (\ref{bloch}) was taken into account
and the contribution $\propto \partial_Q L_B$ in (\ref{spann}) was omitted.) 
For $E_J\ll E_C$ the Bloch inductance $L_B$ is sharply peaked around $Q=e$ (see Fig.~\ref{fig:BlochInductance}). In the opposite case, $E_J\gg E_C$,  the Bloch inductance $L_B$ is nearly 
constant, $L_B\approx L_J\equiv \frac{\Phi_0^2}{4\pi^2 E_J}$ .
Combining Eq.~(\ref{spann}) and the self-consistency equation (\ref{skb1}) we obtain
\begin{eqnarray}
&&L_0\ddot Q_r =-\left[ \frac{2Q_r-Q_{r-1}-Q_{r+1}}{C_0}\right]\nonumber\\
&&-\left[\partial_Q E_0(Q_r)+L_B(Q_r)\ddot Q_r+\frac{1}{2}\partial_Q L_B(Q_r)\dot Q_r^2\right]\ ,\nonumber\\
\label{bwg1}
\end{eqnarray}
where we substituted $\langle Q\rangle\rightarrow Q$ for clarity. We observe that the kinetic inductance is superseded by the Bloch inductance at least around $Q=e$ and we can safely assume $L_0\rightarrow 0$. Yet, in the regime 
$E_J \ll E_C$, when $L_B(Q)$ is exponentially small in the regions $Q \approx 0$ and $Q \approx 2e$, 
a finite geometric or kinetic inductance $L_0$ could be important. In the continuum limit, 
i.e., after the substitution $Q_{r+1}+Q_{r-1}-2Q_r \rightarrow \partial^2_r Q = Q''$, equation (\ref{bwg1}) reads
\begin{equation}
L_B(Q) \ddot Q-\frac{Q''}{C_0}+\frac{1}{2}(\partial_Q L_B)\dot Q^2+\frac{\partial E_0(Q)}{\partial Q}=0\ .
\label{sine1}
\end{equation}
We now make the very important observation that (\ref{sine1}) is the equation of motion for the following Lagrangian density
\begin{equation}
 \mathcal{L}(Q,\dot Q)=\frac{1}{2}L_B(Q)\dot Q^2-\frac{1}{2C_0}Q'^2-E_0(Q)\ .
\label{lagra1}
\end{equation}
In the limit $E_J\gg E_C$, when $L_B\approx const$ and $E_0(Q) \propto \cos Q$ we obtain the 
usual sine-Gordon equation.
On the other hand, in the limit $E_J\ll E_C$ equation (\ref{sine1}) differs in several aspects from the sine-Gordon equation: i) The first two terms of (\ref{sine1}) describe a wave guide with a $Q$-dependent ``light velocity'' $c(Q)=\frac{1}{\sqrt{L_B(Q)C_0}}$. 
With the Bloch inductance having a peak value 
$L_{max}=L_B(e)$ at $Q=e$ we obtain the minimal light velocity $c_{min}=\frac{1}{\sqrt{L_{max}C_0}}$. 
ii) The ground state energy $E_0$ is still a $2e$ periodic function of $Q$ but it is no longer 
proportional to $\cos Q$. iii) Since $L_B$ depends strongly on $Q$, the third term of (\ref{sine1})
is very important. 

\subsection{Solitonic solutions}

We are now searching for a  solitary wave traveling with velocity $v$ by
plugging the ansatz $Q(r-vt)$ into Eq.~(\ref{sine1}). This gives the 
following differential equation
\begin{equation}
\frac{\partial}{\partial r}\left[\frac{1}{2} \left(L_B(Q) v^2 - \frac{1}{C_0}\right) Q'^2+E_0(Q)\right] = 0\ .
\end{equation} 
Integrating we obtain
\begin{equation}
r-r_0=\pm\int\limits_{Q(r_0)}^{Q(r)}dQ\left[\frac{2C_0(E_0(Q)-E_{min})}{1-L_B(Q) C_0v^2}\right]^{-1/2}\ .
\label{xx0}
\end{equation}
Here, $E_{min}$ is an integration constant and $\pm$ stands for the soliton / antisoliton solution. We impose the boundary conditions $Q(-\infty)=0$ and $Q(+\infty)=2e$ to describe the propagation of a single Cooper pair in the array. 
This also fixes the integration constant, $E_{min}=E_0(0)=E_0(2e)$.
The solitonic solutions only exist for $v \le c_{min}$.

\subsection{Lorentz contraction}

\begin{figure}
 \includegraphics[scale=0.2]{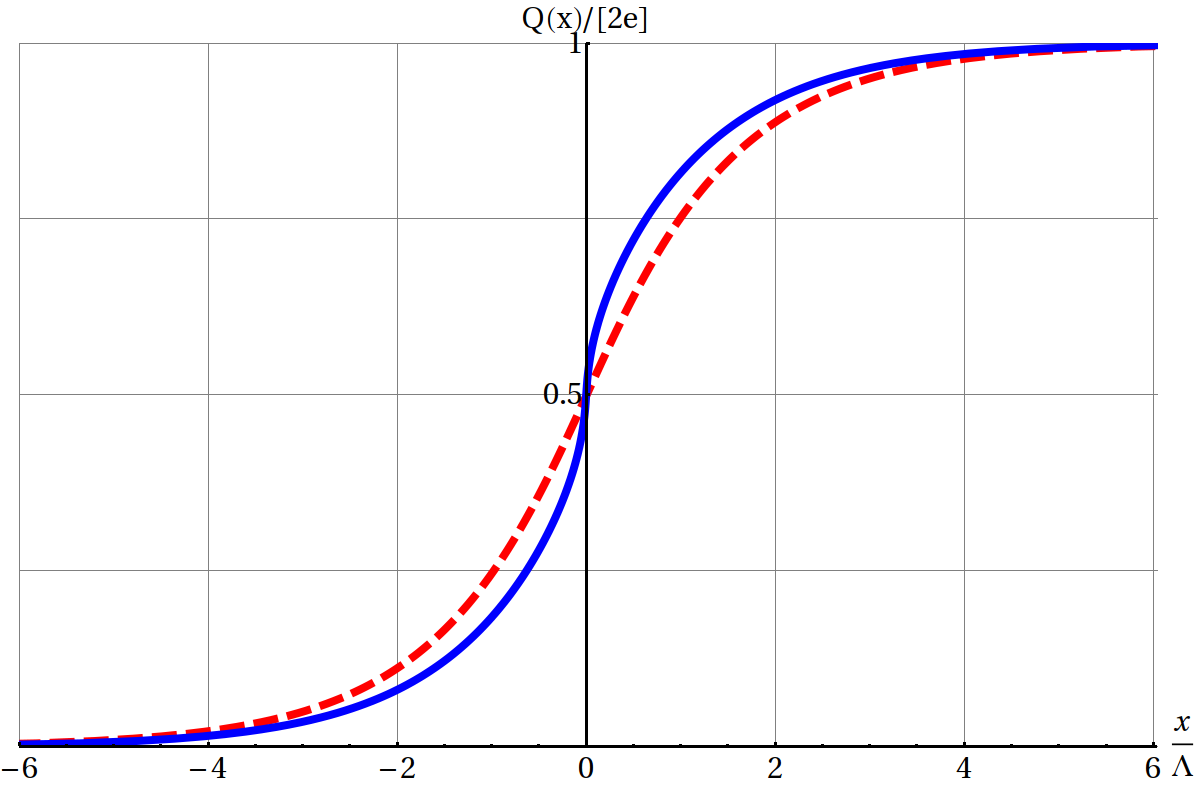}
\caption{The solution $Q(x)$ from (\ref{xx0}) for $v/c_{min}=0.1$ (red dashed curve) and $v/c_{min}=0.99$ (blue solid curve), both for $E_J/E_C=0.5$}
\label{lorentz}
\end{figure}

In the limit $E_J \gg E_C$, Eq.~(\ref{sine1}) reduces to the sine-Gordon equation and is Lorentz invariant. 
Thus solitons undergo the usual Lorentz contraction. In the other limit, $E_J \ll E_C$,  Eq.~(\ref{sine1})  is 
not Lorentz invariant. The Lorentz contraction of the soliton takes a very peculiar shape. 
Consider a soliton moving with velocity $v$ approaching $c_{min}$ (we postpone for a moment 
a discussion on whether this is consistent with adiabaticity). 
For the center of the soliton, where $Q \approx e$, the relativistic regime is reached and it is Lorentz
contracted (see Fig.~\ref{lorentz}). In contrast, the soliton's tales, where $Q \sim 0$ or $Q\sim 2e$
are unaffected by Lorentz contraction.

\subsection{Rest energy and dynamical mass of the soliton}

Using the Lagrangian density (\ref{lagra1}) we find the energy of a soliton
\begin{equation}
E_{sol}(v)=\frac{1}{C_0}\int\limits_{0}^{2e}dQ\sqrt{\frac{2[E_0(Q)-E_0(0)]C_0}{1-v^2L_B(Q)C_0}}\ .
\label{energie1}
\end{equation}
For small velocities we expand and obtain $E_{sol}(v) = E_{rest} + \frac{1}{2}\,m_{kin}v^2 + O(v^4)$.
The rest energy of the soliton is given by 
\begin{equation}
E_{rest}=\int\limits_0^{2e}dQ\sqrt{\frac{2[E_0(Q)-E_0(0)]}{C_0}}\ .
\label{restmass}
\end{equation}
For the kinetic mass we obtain
\begin{equation}
m_{kin}=\int\limits_0^{2e}dQ\,L_B(Q)\,\sqrt{2[E_0(Q)-E_0(0)]C_0}\ .
\label{masse}
\end{equation}
In the limit $E_J \gg E_C$, when $L_B \approx L_J = const.$, we obtain 
as expected the relativistic relation $E_{rest} \approx m^{\phantom 2}_{kin} c_{min}^2$
(in this limit the light velocity is $Q$-independent, $c(Q)\approx c_{min} \approx (L_J C_0)^{-1/2}$).

In the opposite charging limit, $E_J \ll E_C$, no such relativistic relation exists. 
A simple estimate then gives  
\begin{equation}
E_{rest}\approx\Lambda\frac{E_C}{2}\left(1- \mathcal{O}\left[\frac{E_J}{E_C}\right]^2\right), 
\label{ruhenah}
\end{equation}
consistent with the result of Sec.~\ref{subsec:qualit_disc}.

In the limit $E_J\ll E_C$ the Bloch inductance $L_B(Q)$ is sharply peaked around $Q = e$ 
and the integral of Eq.~(\ref{masse}) is dominated by a small vicinity of this point. 
Here a two-state approximation is valid which gives $L_B=2\left(\frac{E_C}{E_J}\right)^2L_J\sin^5(\theta)$ with $\cot\theta\equiv \left(\frac{Q-e}{e}\right)\,\frac{E_C}{E_J}$. As $L_B(Q)$ is sharply peaked around $Q=e$, we can replace $E_0(Q)-E_0(0)$ in (\ref{masse}) by its value at $Q=e$, $\frac{1}{4}E_C-\frac{1}{2}E_J$, leading to
\begin{equation}
 m_{kin}\approx m_{kin}^{bare} \frac{2E_C}{3\Lambda E_J}\,\left(1-\frac{E_J}{E_C}+\mathcal{O}\left[\frac{E_J}{E_C}\right]^2\right)\ ,
\label{massenah}
\end{equation}
where $m_{kin}^{bare} \equiv \hbar^2/E_J$ is the "naive" tight-binding mass of a single Cooper pair. The polaronic reduction of the mass is evident from (\ref{massenah}) in the 
regime $\Lambda E_J > E_C > E_J$.

\subsection{Comparison with the tight-binding results}

\begin{figure}[!htb]
 \includegraphics[width=0.9\columnwidth]{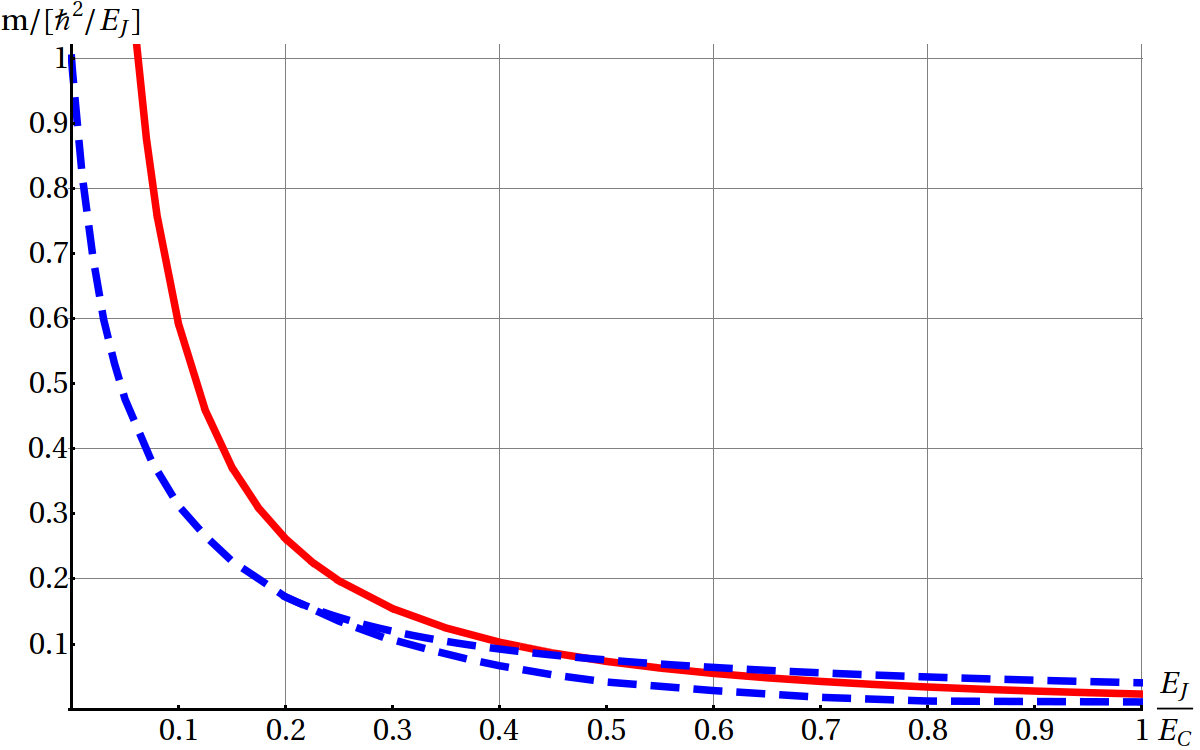}
\caption{Solid (red) curve: mean-field mass $m_{kin}$. Lower dashed (blue) curve: tight-binding mass $m_{TB}$. Upper dashed (blue) curve: upper boundary for $m_{TB}$.}
\label{fig:MassesMFvsTB}
\end{figure}

In Fig.~\ref{fig:MassesMFvsTB} the mass $m_{kin}$ obtained in the mean-field approach 
is compared with the mass $m_{\rm TB}$ from the tight-binding calculation.   
We observe a good correspondence for $E_J/E_C > 0.3$. This is the main result of this paper. 
The mass scales approximately as $\sim E_J^{-2}$. 
As the convergence of the tight-binding approach gets worse with raising ratio $\frac{E_J}{E_C}$ we also show the uncertainty of the result by giving the upper boundary for $m_{TB}$ (upper dashed curve in Fig. \ref{fig:MassesMFvsTB}).

\subsection{Adiabaticity condition and the validity of the mean field theory}
The analysis above rests on two assumptions: a) the dynamics of $Q$ is slow and allows us to neglect the Landau-Zener tunneling in the derivation of Eq.~(\ref{spann}); b) the field $Q$ can be regarded as classical.

Since in terms of $Q$ the solitons are large objects (with the size of the order of $2\Lambda$) one can expect the second assumption  to hold in a wide parameter range. In particular, at $E_J \sim E_C$ we can estimate the ``effective Planck constant''~\cite{PhysRevD.11.2088} for the Lagrangian (\ref{lagra1}) as
\begin{equation}
\beta=\frac{2\pi \hbar}{(2e)^2}\sqrt{\frac{C_0}{L_B}}\sim \frac{1}{\Lambda}\sqrt{\frac{E_J}{E_C}}\ll 1\ .
\end{equation}
While at small enough $E_J/E_C$ strong suppression of the nonlinear Bloch inductance $L_B(Q)$ for $Q\neq e$ may become important, we expect this effect to be of minor significance in the intermidiate range of $E_J/E_C$.

The situation with the assumption a) is much more tricky. First of all, the adiabaticity imposes an upper boundary 
for the velocity of solitons to be considered. In the limit $E_J\ll E_C$ the probability of a Landau-Zener transition into the first excited level is given by 
\begin{equation}
P = e^{-\frac{\pi}{2\hbar}\Delta^2\frac{1}{|\dot\varepsilon|}} \ ,
\label{landau1}
\end{equation}
where $\Delta\approx E_J$, $\varepsilon(Q)$ is the difference of the charging energies of the two charge states involved in the process. Thus, $\dot \varepsilon =\dot Q \partial_Q\varepsilon$ and $|\partial_Q\varepsilon|\approx\frac{E_C}{2e}$.
If we demand $P \le e^{-x}$, the corresponding limitation on the soliton's 
velocity reads
\begin{equation}
 v^2\le c_{min}^2\frac{1}{1+\left(\frac{2x}{\pi}\right)^2\frac{E_C}{E_J}}\ .
\label{landau9}
\end{equation}
We see that for $E_J/E_C \rightarrow 0$, the maximal velocity for which adiabaticity 
still holds goes to zero. 

It is obvious that even for a static soliton the adiabaticity condition can be broken by fluctuations around the saddle point. Thus, the precise determination of the applicability region for the adiabatic approximation requires 
understanding of the characteristic time scale for the two-point correlation function of $Q$ with the account of the nonlinear Bloch inductance. However, the good agreement between the mean-field theory and the tight-binding approach found in the calculation of the soliton mass for intermediate $E_J/E_C\sim 0.5$ allows us to expect that 
adiabaticity indeed holds in this parameter range for small soliton velocities.

\section{Conclusions}

In this paper we studied the dynamical properties of the charge carriers (charge solitons) 
in infinite one-dimensional Josephson arrays without disorder. We applied two complementary techniques and arrived at our main result: in the parameter 
regime $E_J < E_C < \Lambda E_J$ the polaronic effects strongly reduce the 
effective mass of the charge solitons which scales approximately as $E_J^{-2}$. 

\section{Acknowledgements}

We thank M. Berry, A. Ustinov, H. Rotzinger, R. Sch\"afer for numerous 
fruitful discussions. IP acknowledges support from the German-Israeli Foundation (GIF). 
SR acknowledges support from the Deutsche Forschungsgmeinschaft (DFG) 
under Grant No. RA 1949/1-1.

\appendix
\section{Low lying excitations and the spin formulation}
\label{App:Low_energy_space}
The aim of the present Appendix is to find an explicit description of the low energy charge configurations satisfying the constraint
\begin{equation}
	\sum_{rr'}|r-r'|n_{r}n_{r'}=0\ .
\end{equation}
Let us consider one such configuration. We try to add  a dipole at islands $R$ and $R+1$ to this configuration, i.e. we construct a new configuration given by
\begin{equation}
	\tilde{n}_r=n_r+\delta_{rR}-\delta_{r, R+1}\ .
\end{equation}
For the new configuration to belong to the low lying sector we need the following condition to hold:
\begin{equation}
	\sum_{rr'}|r-r'|\tilde{n}_{r}\tilde{n}_{r'}=
	2\sum_{r} n_r\left( |r-R|-|r-R-1|\right) -2 =0\ .
\end{equation}
We can rewrite this condition as
\begin{equation}
	\sum_{r\geq R+1}n_r-\sum_{r\leq R}n_r=1\ .
	\label{cond_sigma}
\end{equation}
In terms of the spin variable $\sigma_r^z$ introduced in Sec. ~\ref{sec:Structure}, Eq. (\ref{cond_sigma}) is equivalent to the condition
\begin{equation}
	\sigma_R^z=-1\ .
\end{equation}
After the creation of an additional dipole (i.e. the Cooper pair tunneling from island $R+1$ to island $R$) the new state is described by  
\begin{equation}
	\tilde{\sigma}^z_r=-\sum_{r'\geq r+1}\tilde{n}_{r'}+\sum_{r'\leq r}\tilde{n}_{r'}=\sigma_r^z+2\delta_{r, R}\ .
\end{equation}
We thus conclude that the 
Cooper pair tunneling from island $R+1$ to island $R$ is allowed (i.e. drives the system into another state within the low energy subspace) if 
$\sigma^z_R=-1$. Such a tunneling corresponds to the spin flip at the link connecting the island $R$ and $R+1$. 
It is easy to check that the inverse process (tunneling from $R$ to $R+1$) is allowed only when $\sigma_R^z=1$ and also leads to the flip of $\sigma_R^z$.
Taking into account that a single Cooper pair at island $R$ is described in terms of $\sigma_r^z$ by a domain wall 
\begin{equation}
	\sigma^z_{r<R}=-1\quad,\quad \sigma^z_{r\leq R}=1\ ,
\end{equation}
we conclude that the low energy configurations are those with $\sigma_r^z=\pm 1$ for all $r$ 
Translated into the charge language this condition gives the conditions mentioned in the main text (Sec.~\ref{sec:Structure}).

\section{Projecting the Hamiltonian}
\label{App:Projection}

We consider the action of the Josephson term in the Hamiltonian (\ref{Eq:H_spin}) on the state (\ref{Eq:basic_state}). Obviously, we can drop all the terms with 
$r>R+w_0$ or $r<R-w_0-1$ from the sum over $r$ since acting on the state (\ref{Eq:basic_state}) they inevitably 
create the configuration of the width greater than $w_0$. The terms $\sigma_r^x$ with $R+1\leq r\leq R+w_0$ do not change the position of the first spin 
up in the chain and, thus, the  coordinate of the soliton $R$.  Thus, their action is described by
\begin{equation}
	\sum_{r=R+1}^{R+w_0}\sigma_r^x|R\rangle|\tilde{\sigma}_1\,, \ldots \tilde{\sigma}_{w_0}\rangle=|R\rangle \sum_{i=1}^{w_0}\tilde{\sigma}_i^x
	|\tilde{\sigma}_1\,, \ldots \tilde{\sigma}_{w_0}\rangle \ .
\end{equation}
We consider now the action of $\sigma_R^x$. It is convenient to introduce the operator of the right cyclic shift $T$ acting on the states 
$|\tilde{\sigma}_1\,, \ldots \tilde{\sigma}_{w_0}\rangle$ according to
\begin{equation}
	T|\tilde{\sigma}_1\,, \ldots \tilde{\sigma}_{w_0}\rangle=|\tilde{\sigma}_{w_0}\,, \tilde{\sigma}_{1}\,,\ldots \tilde{\sigma}_2\rangle\ .
\end{equation}
Assume that in the state $|\tilde{\sigma}_1\,, \ldots \tilde{\sigma}_{w_0}\rangle$ exactly $k$ first spins are down ($-1$) (we require now $0\leq k\leq w_0-1$; 
the case of $w_0$ spins down will be considered separately).
The direction of other spins is arbitrary. In this case the action of $\sigma^x_R$ on our state is given by 
\begin{eqnarray}
	&&\sigma_R^x|R\rangle|\tilde{\sigma}_1\,, \ldots \tilde{\sigma}_{w_0}\rangle\nonumber\\&&=
	|R+k+1\rangle \left[T^{+}\right]^{k+1}\left[\prod_{j=1}^{k+1}\tilde{\sigma}_j^+\right]\tilde{\sigma}_{k+1}^-
	|\tilde{\sigma}_1\,, \ldots \tilde{\sigma}_{w_0}\rangle\ .\nonumber\\
\end{eqnarray}
On the other hand, if in the given state exactly $m\neq k$ first  spins are $-1$, then
\begin{equation}
	\left[\prod_{j=1}^{k+1}\tilde{\sigma}_j^+\right]\tilde{\sigma}_{k+1}^-
	|\tilde{\sigma}_1\,, \ldots \tilde{\sigma}_{w_0}\rangle=0\ .
\end{equation}
Thus we conclude that for any state (except the state with all spins down) 
\begin{eqnarray}
	&&\sigma_R^x|R\rangle|\tilde{\sigma}_1\,, \ldots \tilde{\sigma}_{w_0}\rangle\nonumber\\&&=
	\sum_{k=0}^{w_0-1}|R+k+1\rangle \left[T^{+}\right]^{k+1}\left[\prod_{j=1}^{k+1}\tilde{\sigma}_j^+\right]\tilde{\sigma}_{k+1}^-
	|\tilde{\sigma}_1\,, \ldots \tilde{\sigma}_{w_0}\rangle\ .\nonumber\\
\end{eqnarray}
Finally, taking into account that
\begin{equation}
	\sigma_R^x|R\rangle|\downarrow\,,\ldots \downarrow\rangle=|R+w_0+1\rangle \left[\prod_{j=1}^{w_0}\tilde{\sigma}_j^+\right]
	|\downarrow\,,\ldots \downarrow\rangle
\end{equation}
we find for the projection of $\sigma_{R}^x$ onto the space spaned by the configurations of the width less or equal than $w_0$ 
\begin{eqnarray}
	&&\sigma_R^x|R\rangle|\tilde{\sigma}_1\,, \ldots \tilde{\sigma}_{w_0}\rangle\nonumber\\&&=
	\left[\sum_{k=1}^{w_0}
	|R+k\rangle\left[T^{+}\right]^{k}\left[\prod_{j=1}^{k}\tilde{\sigma}_j^+\right]\tilde{\sigma}_{k}^- \right.\nonumber\\&&\left.
	+|R+w_0+1\rangle \left[\prod_{j=1}^{w_0}\tilde{\sigma}_j^+\right]\right]
	|\tilde{\sigma}_1\,, \ldots \tilde{\sigma}_{w_0}\rangle\ .
	\label{Eq:sigma_R}
\end{eqnarray}
The last part of the Josephson Hamiltonian 
\begin{equation}
	\sum_{r=R-1}^{R-w-1}\sigma_r^x \ ,
\end{equation}
after the projection onto the subspace of interest, produces an expression conjugate to (\ref{Eq:sigma_R}). 
Thus, summing up all the contributions we find the projected Josephson Hamiltonian
\begin{multline}
	H_J|R\rangle|\tilde{\sigma}_1\,, \ldots \tilde{\sigma}_{w_0}\rangle=
	-E_J |R\rangle \sum_{k=1}^{w_0}\tilde{\sigma}_k^x
	|\tilde{\sigma}_1\,, \ldots \tilde{\sigma}_{w_0}\rangle\\
	-E_J\left[\sum_{k=1}^{w_0}
	|R+k\rangle\left[T^{+}\right]^{k}\left[\prod_{j=1}^{k}\tilde{\sigma}_j^+\right]\tilde{\sigma}_{k}^-\right.\\
	\left.+|R+w_0+1\rangle \left[\prod_{j=1}^{w_0}\tilde{\sigma}_j^+\right]
	+h.c.
	\right]
	|\tilde{\sigma}_1\,, \ldots \tilde{\sigma}_{w_0}\rangle\ .
\end{multline}
Going to the momentum domain with respect to the cyclic coordinate $R$ and performing some simplifications based on the elementary properties 
of $T$ and $\tilde{\sigma}^{\pm}_k$, we finally arrive at Eq.~(\ref{Eq:H_final}).

\section{Adiabatic calculation}
\label{App:Adiabatic}

The most direct way is to apply the time-dependent perturbation theory. We will instead use a technique proposed by Berry~\cite{ProcRSocLondA.414.31} where we apply several unitary transformations, so that the eigenstates of the transformed Hamiltonians approach asymptotically the actual evolving state. 

We consider the adiabatic case and the terms of orders higher than $\ddot Q, \dot Q^2$ are omitted. The general transformation reads
\begin{equation}
  H_{k+1}(t)= U_k^{\dagger}H_kU_k-i\hbar U_k^{\dagger}\dot U_k\ ,
\label{berrytrafo1}
\end{equation}
with $U_k(t)$ being the time-dependent unitary operator that diagonalizes $H_k$ at each time $t$. We can write
\begin{equation}
 \Ket{e_n^{(k)}(t)}=U_k(t)\Ket{n}\ ,
\label{berrytrafo2}
\end{equation}
where $ \Ket{e_n^{(k)}(t)}$ are the eigenvectors satisfying 
\begin{equation}
H_k(t)\Ket{e_n^{(k)}(t)}=E_n^{(k)}(t)\Ket{e_n^{(k)}(t)}\ ,
\label{berrytrafo3}
\end{equation}
and time-independent vectors $\Ket{n}$ can be chosen arbitrarily, e.g., $\Ket{n}=\Ket{e_n^{(0)}(t=-\infty)}$. We now perform the first step of this process explicitly. 
As Hamiltonian $H_0(t)$ we take (\ref{mfnham2}) with $\Ket{e_n^{(0)}(t)}=\Ket{e_n(Q(t))}$.
From (\ref{berrytrafo2}) we find
\begin{equation}
 U_0(t)=\sum\limits_n \Ket{e_n^{(0)}(t)}\Bra{n}\ .
\label{unitrafo}
\end{equation}
Applying (\ref{berrytrafo1}) gives the transformed Hamiltonian
\begin{equation}
H_1(t)=\sum\limits_n\Ket{n}\Bra{n}E_n^{(0)}-i\hbar\sum\limits_{\substack{m,n\\m\not=n}}\Ket{n}\Braket{e_n^{(0)}|\dot e_m^{(0)}}\Bra{m}\ .
\label{berry2}
\end{equation}
We find $\Ket{e_n^{(1)}(t)}$ by applying the usual time-independent perturbation theory with the perturbation being the second term on the RHS of (\ref{berry2}). This gives the new transformation matrix $U_1(t)$ because $\Ket{e_n^{(1)}(t)}=U_1(t)\Ket{n}$, which can be used for a second transformation to obtain $H_2(t)$ with eigenvectors $\Ket{e_n^{(2)}(t)}$.
After calculating $\Ket{e_n^{(2)}(t)}$ we can go back to the desired $\Ket{e_n^{(0)}(t)}\equiv\Ket{e_n}$ basis via
\begin{equation}
\Ket{\Psi(t)}=U_0(t)U_1(t)\Ket{e_n^{(2)}(t)}\ .
\label{trafoback}
\end{equation}
Putting everything together gives
\begin{widetext}
\begin{equation}
  \Ket{\Psi(t)}=\Ket{e_n}\left[1-\frac{1}{2}\sum\limits_{m\not=n}\frac{f_{mn}^2(t)}{\Omega_{mn}^2(t)}\right]+\sum\limits_{m\not=n}\Ket{e_m}
  \left[\frac{f_{mn}}{i\Omega_{mn}}-\sum\limits_{k\not=n}\frac{f_{mk}f_{kn}}{\Omega_{mn}\Omega_{kn}}+\frac{\dot f_{mn}}{\Omega_{mn}^2}-\frac{f_{mn}}{\Omega_{mn}^3}\frac{\partial\Omega_{mn}}{\partial Q}\dot Q\right]\ ,
\label{psi2}
\end{equation}
\end{widetext}
with $f_{nm}\equiv\dot Q \Braket{\partial_Q e_n|e_m}$ and $\Omega_{nm}\equiv\frac{E_n-E_m}{\hbar}$.
\newline
We are now able to calculate the voltage $V_r$ with $\Ket{\Psi(t)}$, where we set $n=0$ in (\ref{psi2}) as we consider our system being initially in the ground state. We use $\Braket{e_n|\partial_Q H|e_m}=\hbar\Omega_{nm}\Braket{\partial_Q e_n|e_m}\text{ for }n\not=m$ and $\Braket{e_n|\partial_Q H|e_n}=\partial_Q E_n$ and arrive 
at Eqs.~(\ref{spann},\ref{bloch}).

\bibliography{charge_soliton}

\begin{thebibliography}{21}
\expandafter\ifx\csname natexlab\endcsname\relax\def\natexlab#1{#1}\fi
\expandafter\ifx\csname bibnamefont\endcsname\relax
  \def\bibnamefont#1{#1}\fi
\expandafter\ifx\csname bibfnamefont\endcsname\relax
  \def\bibfnamefont#1{#1}\fi
\expandafter\ifx\csname citenamefont\endcsname\relax
  \def\citenamefont#1{#1}\fi
\expandafter\ifx\csname url\endcsname\relax
  \def\url#1{\texttt{#1}}\fi
\expandafter\ifx\csname urlprefix\endcsname\relax\def\urlprefix{URL }\fi
\providecommand{\bibinfo}[2]{#2}
\providecommand{\eprint}[2][]{\url{#2}}

\bibitem[{\citenamefont{Bradley and Doniach}(1984)}]{PhysRevB.30.1138}
\bibinfo{author}{\bibfnamefont{R.~M.} \bibnamefont{Bradley}} \bibnamefont{and}
  \bibinfo{author}{\bibfnamefont{S.}~\bibnamefont{Doniach}},
  \bibinfo{journal}{Phys. Rev. B} \textbf{\bibinfo{volume}{30}},
  \bibinfo{pages}{1138} (\bibinfo{year}{1984}).

\bibitem[{\citenamefont{Odintsov}(1994)}]{JETPLett.60.738}
\bibinfo{author}{\bibfnamefont{A.~A.} \bibnamefont{Odintsov}},
  \bibinfo{journal}{JETP Lett.} \textbf{\bibinfo{volume}{60}},
  \bibinfo{pages}{738} (\bibinfo{year}{1994}).

\bibitem[{\citenamefont{Odintsov}(1996)}]{PhysRevB.54.1228}
\bibinfo{author}{\bibfnamefont{A.~A.} \bibnamefont{Odintsov}},
  \bibinfo{journal}{Phys. Rev. B} \textbf{\bibinfo{volume}{54}},
  \bibinfo{pages}{1228} (\bibinfo{year}{1996}).

\bibitem[{\citenamefont{Haviland and Delsing}(1996)}]{PhysRevB.54.R6857}
\bibinfo{author}{\bibfnamefont{D.~B.} \bibnamefont{Haviland}} \bibnamefont{and}
  \bibinfo{author}{\bibfnamefont{P.}~\bibnamefont{Delsing}},
  \bibinfo{journal}{Phys. Rev. B} \textbf{\bibinfo{volume}{54}},
  \bibinfo{pages}{R6857} (\bibinfo{year}{1996}).

\bibitem[{\citenamefont{Hermon et~al.}(1996)\citenamefont{Hermon, Ben-Jacob,
  and Sch\"on}}]{PhysRevB.54.1234}
\bibinfo{author}{\bibfnamefont{Z.}~\bibnamefont{Hermon}},
  \bibinfo{author}{\bibfnamefont{E.}~\bibnamefont{Ben-Jacob}},
  \bibnamefont{and} \bibinfo{author}{\bibfnamefont{G.}~\bibnamefont{Sch\"on}},
  \bibinfo{journal}{Phys. Rev. B} \textbf{\bibinfo{volume}{54}},
  \bibinfo{pages}{1234} (\bibinfo{year}{1996}).

\bibitem[{\citenamefont{Glazman and Larkin}(1997)}]{PhysRevLett.79.3736}
\bibinfo{author}{\bibfnamefont{L.~I.} \bibnamefont{Glazman}} \bibnamefont{and}
  \bibinfo{author}{\bibfnamefont{A.~I.} \bibnamefont{Larkin}},
  \bibinfo{journal}{Phys. Rev. Lett.} \textbf{\bibinfo{volume}{79}},
  \bibinfo{pages}{3736} (\bibinfo{year}{1997}).

\bibitem[{\citenamefont{Haviland et~al.}(2000)\citenamefont{Haviland,
  Andersson, and {\AA}gren}}]{JLTP.118.733}
\bibinfo{author}{\bibfnamefont{D.~B.} \bibnamefont{Haviland}},
  \bibinfo{author}{\bibfnamefont{K.}~\bibnamefont{Andersson}},
  \bibnamefont{and}
  \bibinfo{author}{\bibfnamefont{P.}~\bibnamefont{{\AA}gren}},
  \bibinfo{journal}{J. of Low Temp. Phys.} \textbf{\bibinfo{volume}{118}},
  \bibinfo{pages}{733} (\bibinfo{year}{2000}).

\bibitem[{\citenamefont{{\AA}gren et~al.}(2001)\citenamefont{{\AA}gren,
  Andersson, and Haviland}}]{JLTP.124.291}
\bibinfo{author}{\bibfnamefont{P.}~\bibnamefont{{\AA}gren}},
  \bibinfo{author}{\bibfnamefont{K.}~\bibnamefont{Andersson}},
  \bibnamefont{and} \bibinfo{author}{\bibfnamefont{D.~B.}
  \bibnamefont{Haviland}}, \bibinfo{journal}{J. of Low Temp. Phys.}
  \textbf{\bibinfo{volume}{124}}, \bibinfo{pages}{291} (\bibinfo{year}{2001}).

\bibitem[{\citenamefont{Gurarie and Tsvelik}(2004)}]{JLTP.135.245}
\bibinfo{author}{\bibfnamefont{V.}~\bibnamefont{Gurarie}} \bibnamefont{and}
  \bibinfo{author}{\bibfnamefont{A.}~\bibnamefont{Tsvelik}},
  \bibinfo{journal}{J. of Low Temp. Phys.} \textbf{\bibinfo{volume}{135}},
  \bibinfo{pages}{245} (\bibinfo{year}{2004}).

\bibitem[{\citenamefont{Efetov}(1980)}]{SovJETP.51.1015}
\bibinfo{author}{\bibfnamefont{K.~B.} \bibnamefont{Efetov}},
  \bibinfo{journal}{Sov. Phys. JETP} \textbf{\bibinfo{volume}{51}},
  \bibinfo{pages}{1015} (\bibinfo{year}{1980}).

\bibitem[{\citenamefont{Mooij et~al.}(1990)\citenamefont{Mooij, van Wees,
  Geerligs, Peters, Fazio, and Sch\"on}}]{PhysRevLett.65.645}
\bibinfo{author}{\bibfnamefont{J.~E.} \bibnamefont{Mooij}},
  \bibinfo{author}{\bibfnamefont{B.~J.} \bibnamefont{van Wees}},
  \bibinfo{author}{\bibfnamefont{L.~J.} \bibnamefont{Geerligs}},
  \bibinfo{author}{\bibfnamefont{M.}~\bibnamefont{Peters}},
  \bibinfo{author}{\bibfnamefont{R.}~\bibnamefont{Fazio}}, \bibnamefont{and}
  \bibinfo{author}{\bibfnamefont{G.}~\bibnamefont{Sch\"on}},
  \bibinfo{journal}{Phys. Rev. Lett.} \textbf{\bibinfo{volume}{65}},
  \bibinfo{pages}{645} (\bibinfo{year}{1990}).

\bibitem[{\citenamefont{Fisher}(1990)}]{PhysRevLett.65.923}
\bibinfo{author}{\bibfnamefont{M.~P.~A.} \bibnamefont{Fisher}},
  \bibinfo{journal}{Phys. Rev. Lett.} \textbf{\bibinfo{volume}{65}},
  \bibinfo{pages}{923} (\bibinfo{year}{1990}).

\bibitem[{\citenamefont{Fazio and Sch\"on}(1991)}]{PhysRevB.43.5307}
\bibinfo{author}{\bibfnamefont{R.}~\bibnamefont{Fazio}} \bibnamefont{and}
  \bibinfo{author}{\bibfnamefont{G.}~\bibnamefont{Sch\"on}},
  \bibinfo{journal}{Phys. Rev. B} \textbf{\bibinfo{volume}{43}},
  \bibinfo{pages}{5307} (\bibinfo{year}{1991}).

\bibitem[{\citenamefont{Fazio and {van der Zant}}(2001)}]{PhysRep.355.235}
\bibinfo{author}{\bibfnamefont{R.}~\bibnamefont{Fazio}} \bibnamefont{and}
  \bibinfo{author}{\bibfnamefont{H.}~\bibnamefont{{van der Zant}}},
  \bibinfo{journal}{Phys. Rep.} \textbf{\bibinfo{volume}{355}},
  \bibinfo{pages}{235} (\bibinfo{year}{2001}).

\bibitem[{\citenamefont{Zorin}(2006)}]{PhysRevLett.96.167001}
\bibinfo{author}{\bibfnamefont{A.~B.} \bibnamefont{Zorin}},
  \bibinfo{journal}{Phys. Rev. Lett.} \textbf{\bibinfo{volume}{96}},
  \bibinfo{pages}{167001} (\bibinfo{year}{2006}).

\bibitem[{\citenamefont{Hutter et~al.}(2006)\citenamefont{Hutter, Shnirman,
  Makhlin, and Sch{\"o}n}}]{EPL.74.1088}
\bibinfo{author}{\bibfnamefont{C.}~\bibnamefont{Hutter}},
  \bibinfo{author}{\bibfnamefont{A.}~\bibnamefont{Shnirman}},
  \bibinfo{author}{\bibfnamefont{Y.}~\bibnamefont{Makhlin}}, \bibnamefont{and}
  \bibinfo{author}{\bibfnamefont{G.}~\bibnamefont{Sch{\"o}n}},
  \bibinfo{journal}{Europhys. Lett.} \textbf{\bibinfo{volume}{74}},
  \bibinfo{pages}{1088} (\bibinfo{year}{2006}).

\bibitem[{\citenamefont{Rachel and Shnirman}(2009)}]{PhysRevB.80.180508}
\bibinfo{author}{\bibfnamefont{S.}~\bibnamefont{Rachel}} \bibnamefont{and}
  \bibinfo{author}{\bibfnamefont{A.}~\bibnamefont{Shnirman}},
  \bibinfo{journal}{Phys. Rev. B} \textbf{\bibinfo{volume}{80}},
  \bibinfo{pages}{180508} (\bibinfo{year}{2009}).

\bibitem[{\citenamefont{Zhang et~al.}(2005)\citenamefont{Zhang, Kamenev, and
  Shklovskii}}]{PhysRevLett.95.148101}
\bibinfo{author}{\bibfnamefont{J.}~\bibnamefont{Zhang}},
  \bibinfo{author}{\bibfnamefont{A.}~\bibnamefont{Kamenev}}, \bibnamefont{and}
  \bibinfo{author}{\bibfnamefont{B.~I.} \bibnamefont{Shklovskii}},
  \bibinfo{journal}{Phys. Rev. Lett.} \textbf{\bibinfo{volume}{95}},
  \bibinfo{pages}{148101} (\bibinfo{year}{2005}).

\bibitem[{\citenamefont{Bon\ifmmode~\check{c}\else \v{c}\fi{}a
  et~al.}(1999)\citenamefont{Bon\ifmmode~\check{c}\else \v{c}\fi{}a, Trugman,
  and Batisti\ifmmode~\acute{c}\else \'{c}\fi{}}}]{PhysRevB.60.1633}
\bibinfo{author}{\bibfnamefont{J.}~\bibnamefont{Bon\ifmmode~\check{c}\else
  \v{c}\fi{}a}}, \bibinfo{author}{\bibfnamefont{S.~A.} \bibnamefont{Trugman}},
  \bibnamefont{and}
  \bibinfo{author}{\bibfnamefont{I.}~\bibnamefont{Batisti\ifmmode~\acute{c}\el%
se \'{c}\fi{}}}, \bibinfo{journal}{Phys. Rev. B} \textbf{\bibinfo{volume}{60}},
  \bibinfo{pages}{1633} (\bibinfo{year}{1999}).

\bibitem[{\citenamefont{Coleman}(1975)}]{PhysRevD.11.2088}
\bibinfo{author}{\bibfnamefont{S.}~\bibnamefont{Coleman}},
  \bibinfo{journal}{Phys. Rev. D} \textbf{\bibinfo{volume}{11}},
  \bibinfo{pages}{2088} (\bibinfo{year}{1975}).

\bibitem[{\citenamefont{Berry}(1987)}]{ProcRSocLondA.414.31}
\bibinfo{author}{\bibfnamefont{M.~V.} \bibnamefont{Berry}},
  \bibinfo{journal}{Proc. R. Soc. Lond. A} \textbf{\bibinfo{volume}{414}},
  \bibinfo{pages}{31} (\bibinfo{year}{1987}).

\end{thebibliography}

\end{document}